\documentclass[aps,prl,twocolumn,groupedaddress]{revtex4-1}
\usepackage[latin9]{inputenc}
\usepackage{color}
\usepackage{amsmath}
\usepackage{amssymb}
\usepackage{graphicx}

\makeatletter
 
 \@ifundefined{textcolor}{}
 {%
   \definecolor{BLACK}{gray}{0}
   \definecolor{WHITE}{gray}{1}
   \definecolor{RED}{rgb}{1,0,0}
   \definecolor{GREEN}{rgb}{0,1,0}
   \definecolor{BLUE}{rgb}{0,0,1}
   \definecolor{CYAN}{cmyk}{1,0,0,0}
   \definecolor{MAGENTA}{cmyk}{0,1,0,0}
   \definecolor{YELLOW}{cmyk}{0,0,1,0}
 }

\makeatother

\usepackage[english]{babel}

\begin{document}

\title{Sympathetic EIT laser cooling of motional modes in an ion chain}

\author{Y. Lin}

\email{Electronic address: yiheng.lin@nist.gov}

\author{J. P. Gaebler, T. R. Tan, R. Bowler, J. D. Jost}

\email{Current address: École Polytechnique Fédérale de Lausanne, Lausanne, Switzerland}

\author{D. Leibfried}

\author{D. J. Wineland}

\address{\textcolor{black}{National Institute of Standards and Technology,
325 Broadway, Boulder, CO 80305, USA}}
\begin{abstract}
We use electromagnetically induced transparency (EIT) laser cooling
to cool motional modes of a linear ion chain. As a demonstration,
we apply EIT cooling on $^{24}$Mg$^{+}$ ions to cool the axial modes
of a $^{9}$Be$^{+}$-$^{24}$Mg$^{+}$ ion pair and a $^{9}$Be$^{+}$-$^{24}$Mg$^{+}$-$^{24}$Mg$^{+}$-$^{9}$Be$^{+}$
ion chain, thereby sympathetically cooling the $^{9}$Be$^{+}$ ions.
\textcolor{black}{Compared to previous implementations of conventional
Raman sideband cooling, we achieve approximately an order-of-magnitude
reduction in the duration required to cool the modes to near the ground
state and significant reduction in required laser intensity.}
\end{abstract}
\maketitle
One proposal for building a quantum information processor is to use
trapped, laser-cooled ions \cite{CandZ,Blatt2008,Blatt2012}, where
internal states of the ions serve as individual qubits that are manipulated
by laser beams and/or microwave radiation. The Coulomb coupling between
ions establishes normal modes of motion; transitions involving both
the qubit states and motional modes enable entangling gate operations
between multiple qubits. For high-fidelity deterministic entangling
gates, we require that the thermal or uncontrolled components of the
relevant modes be in the Lamb-Dicke regime \cite{Blatt2008}, where
the amplitude of the ions' uncontrolled motion is much less than the
effective wavelength of the coupling radiation \cite{Note1}. For most experiments this means that the motion must be cooled to
near the quantum-mechanical ground state, which has typically been
achieved with sideband laser cooling \textcolor{black}{\cite{Diedrich1989,Monroe1995,Blatt2008}}.
Scaling can potentially be achieved by storing ions in multi-zone
arrays where information is moved in the processor by physically transporting
the ions \cite{Wineland1998,kielpinski02} or teleporting \cite{Gottesman1999}.

Ion motion can be excited by ambient noisy electric fields and/or
during \textcolor{black}{ion} transport \textcolor{black}{\cite{Wineland1998}.}
Therefore, for lengthy algorithms, a method for recooling the ions
is needed. This can be accomplished by combining the qubit ions with
\textcolor{black}{``refrigerant''} ions that are cooled without
disturbing the qubit states, but ``sympathetically'' cool the qubits
through Coulomb coupling\textcolor{blue}{{} }\textcolor{black}{\cite{Wineland1998,kielpinski02,schmidt05,Jost09,Home2009,Hanneke2010,Jost2010,Wubbena2012a}}.
Demonstrations of this technique in information processing have so
far used sideband cooling \cite{schmidt05,Jost09,Home2009,Hanneke2010,Jost2010}.
While effective, sideband cooling can typically cool only one mode
at a time, due to the differences in mode frequencies and narrowness
of the sideband transitions. Furthermore, in the case of stimulated-Raman
transition sideband cooling \cite{Monroe1995}, the laser-beam intensities
and detuning must be sufficiently large to avoid heating from spontaneous
emission. Importantly, in experiments performed in this scalable configuration,
the time required for re-cooling has been the limiting factor \cite{Home2009,Gaebler2012} and leads to errors due to qubit dephasing \cite{Langer05}. A technique that can mitigate these problems is EIT laser
cooling, described theoretically in \cite{Morigi2000,Morigi2003}
and demonstrated on a single ion in \textcolor{black}{\cite{Roos2000,Schmidt-Kaler2001,Webster2005}}.
For EIT cooling, required laser intensities are relatively small and
the cooling bandwidth is large enough that multiple modes can be cooled
simultaneously. To demonstrate these features, we investigate EIT
cooling of multiple modes of linear ion chains containing $^{9}$Be$^{+}$
and $^{24}$Mg$^{+}$ ions. EIT cooling is applied to the $^{24}$Mg$^{+}$
ions, which cools all modes along the axis of the chain to near the
ground state, thereby sympathetically cooling the $^{9}$Be$^{+}$
ions. We realize significant reductions in cooling duration and required
laser intensity compared to previous experiments that employed sideband
cooling \cite{Jost09,Home2009,Hanneke2010,Jost2010}.

\begin{figure}
\includegraphics[scale=0.4]{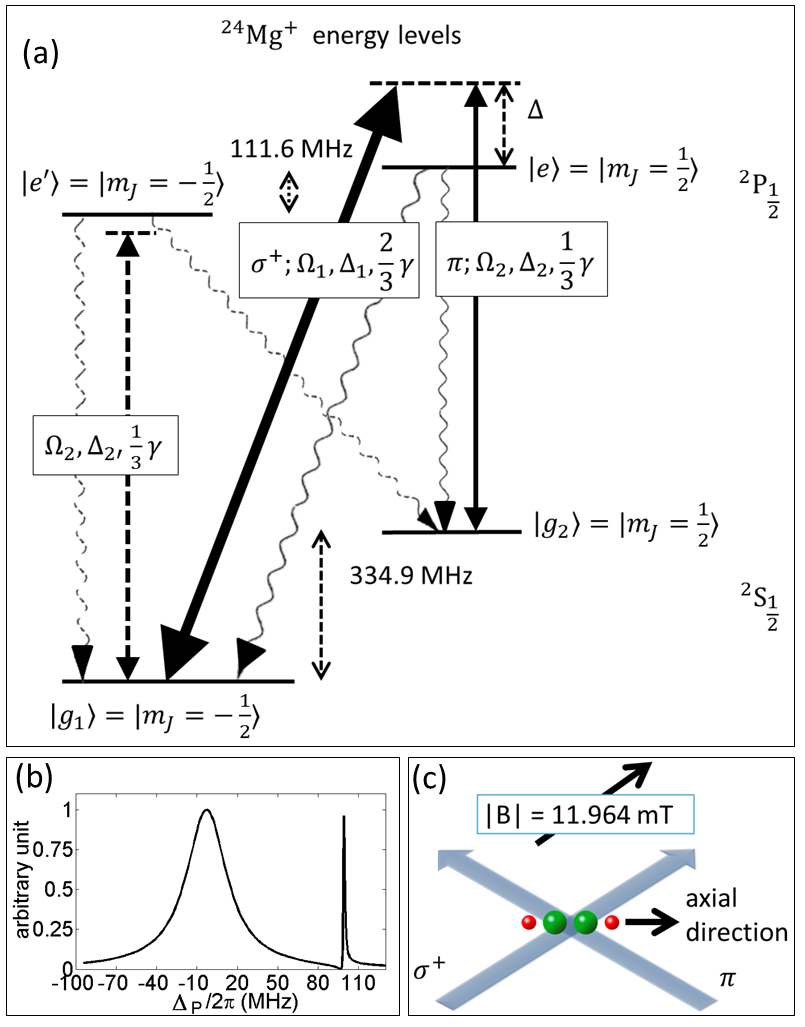} \caption{(a) Relevant energy levels for $^{24}$Mg$^{+}$. The three levels
$|g_{1}\rangle$, $|g_{2}\rangle$ and $|e\rangle$ serve as a $\Lambda$
system for EIT c\textcolor{black}{ooling}. Laser beams with $\sigma^{+}$
and $\pi$ polarizations couple the ground states to the excited state
with Rabi rates $\Omega_{1}$ and $\Omega_{2}$ and detuning $\Delta$.
Wavy lines show spontaneous emission from the excited state to the
ground states and\textcolor{black}{{} the excited-level decay rate is
denoted with $\gamma\simeq$ $2\pi\times$41 MHz. The fourth level
$|e'\rangle$ can perturb} the EIT cooling when the $\pi$ polarized
laser beam has frequency near the $|g_{1}\rangle$ to $|e'\rangle$
resonance. (b) Simulation of the absorption spectrum of a stationary
ion by a weak probe beam for $\Delta$ = $2\pi\times$96.7 MH\textcolor{black}{z,
$\Omega_{1}/2\pi$ = 30 MHz, and $\Omega_{2}/2\pi$ = 12 MHz. For
simplicity, the fourth level }$|e'\rangle$ is ignored for (b). The
probe detuning from the $|g_{2}\rangle$ to $|e\rangle$ resonance
is denoted by $\Delta_{{\rm P}}$. This Fano-like profile contains
a narrow and broad feature corresponding to \textcolor{black}{dressed
states} $|\psi_{+}\rangle$ and $|\psi_{-}\rangle$ respectively \cite{Morigi2003}.
When $\Delta_{\mathrm{P}}=\Delta$, absorption vanishes due to coherent
population trapping. (c) Beam \textcolor{black}{configuration and
a depiction of the $^{9}$Be$^{+}-$$^{24}$Mg$^{+}-{}^{24}$ Mg$^{+}-$$^{9}$Be$^{+}$ion
chain.} \label{fig:Mgstruc_beam_prof}}
\end{figure}

Following \textcolor{black}{\cite{Morigi2003}}, consider the three-level
$\Lambda$ system comprised of the bare states $|g_{1}\rangle,\ |g_{2}\rangle$
and $|e\rangle$ shown in Fig. \ref{fig:Mgstruc_beam_prof}(a). For
an ion at rest, laser beams with resonant Rabi rates $\Omega_{1}$
and $\Omega_{2}$ and equal detunings $\Delta_{1}=\Delta_{2}\equiv\Delta>0$
dress the bare states such that the system relaxes to the ``dark''
steady state $|\psi_{D}\rangle=(\Omega_{2}|g_{1}\rangle-\Omega_{1}|g_{2}\rangle)/\Omega$
with $\Omega\equiv\sqrt{{\Omega_{1}^{2}+\Omega_{2}^{2}}}$. Absorption
from a weak (third) probe laser beam has a spectrum indicated in Fig.
\ref{fig:Mgstruc_beam_prof}(b). The frequency shift between the absorption
null and the relatively narrow peak on the right is
\begin{equation}
\delta=(\sqrt{{\Delta^{2}+\Omega^{2}}}-\Delta)/2.\label{eq:Stark Shift}
\end{equation}
If the difference in k-vectors for the two dressing beams has a component
along the direction of a motional mode, the ion\textquoteright{}s
motion will prevent it from being in the dark state. In the ion\textquoteright{}s
frame of reference, the laser beams appear to be frequency modulated
at the mode frequency $\omega$. For small amplitudes of motion such
that the ion is in the Lamb-Dicke regime, the ion is probed by sidebands
at frequencies $\Delta\pm\omega$. If conditions are such that $\delta\simeq\omega$,
the upper sideband is resonant with the narrow feature on the right
side of Fig. \ref{fig:Mgstruc_beam_prof}(b) and the ion can scatter
a photon while simultaneously losing one quantum of motion, similar
to more conventional sideband cooling. One advantage of this scheme
is that the width of the right-hand peak can be made broad enough
that the condition for cooling is met for multiple modes for the same
value of $\delta$. This may prove advantageous in experiments involving
many ions, such as simulations where the mode frequencies have a relatively
narrow distribution \cite{GDLin2009,Islam2011,Sawyer2012,Islam2012}.

We trap $^{9}$Be$^{+}$ and $^{24}$Mg$^{+}$ ions in a linear radio-frequency
Paul trap described in \cite{Jost2010}, depicted schematically in
Fig. \ref{fig:Mgstruc_beam_prof}(c). The ions form a linear chain
along the axis of the trap, the axis of weakest confinement. We perform
experiments on either a single $^{9}$Be$^{+}$ -$^{24}$Mg$^{+}$
pair or a four-ion chain with the ions in the order $^{9}$Be$^{+}$-$^{24}$Mg$^{+}$-$^{24}$Mg$^{+}$-$^{9}$Be$^{+}$
\cite{Jost09,Jost2010}. A single trapped $^{9}$Be$^{+}$ ion has
motional freque\textcolor{black}{ncy $\omega_{z}/2\pi=2.97$ MHz alo}ng
the trap axis and\textcolor{red}{{} }\textcolor{black}{$\{\omega_{x}/2\pi,\omega_{y}/2\pi\}=\{12.4,11.7\}$}
MHz, along the transverse directions. An internal-state quantization
magnetic field B is applied along a direction $45^{\circ}$ to the
trap axis (Fig. \ref{fig:Mgstruc_beam_prof}(c)), which breaks the
degeneracy of magnetic sublevels of $^{9}$Be$^{+}$ and $^{24}$Mg$^{+}$.
In Fig. \ref{fig:Mgstruc_beam_prof}(a), $m_{J}$ indicates the projection
of the $^{24}$Mg$^{+}$ ion's angular momentum along the direction
of B. For B = 11.964 mT, the energy splitting of the qubit states
\textcolor{black}{$2s\ ^{2}$S$_{1/2}$} $|F=2,m_{F}=1\rangle$ and
$|F=1,m_{F}=0\rangle$ of $^{9}$Be$^{+}$ is first-order insensitive
to changes in B, leading to long coherence ti\textcolor{black}{mes
of superposition states \cite{Langer05}.}

We apply two laser beams near the \textcolor{black}{$3s\ ^{2}$S$_{1/2}$
to $3p\ ^{2}$P$_{1/2}$} transition in $^{24}$Mg$^{+}$ at approximately
280.353 nm (Fig. \ref{fig:Mgstruc_beam_prof}(a)). \textcolor{black}{These
two beams are derived from the same laser and frequency}\textcolor{blue}{{}
}\textcolor{black}{shifted by acousto-optic modulators}\textcolor{blue}{{}
}\textcolor{black}{\cite{Monroe1995a}.} As indicated in Fig. \ref{fig:Mgstruc_beam_prof}(c),
one of the beams propagates along the direction of B with $\sigma^{+}$
polarization to couple $|g_{1}\rangle$ to $|e\rangle$ with resonant
Rabi rate $\Omega_{1}$ and detuning $\Delta_{1}$ from the excited
state. The other beam has $\pi$ polarization and couples $|g_{2}\rangle$
to $|e\rangle$ with resonant Rabi rate $\Omega_{2}$ and detuning
$\Delta_{2}$. We set $\Delta_{1}=\Delta_{2}=\Delta$; ($\Delta/2\pi$
can be set to a precision of approximately 1.5 MHz.) The difference
wave-vector of the two beams is parallel to the trap axis. The values
of $\Omega_{1}$ and $\Omega_{2}$ are determined from measurements
of the Rabi rate for Raman carrier transitions and the AC Stark shift
from the $\sigma^{+}$ polarized beam when it is detuned from resonance.

We first apply Doppler cooling to $^{9}$Be$^{+}$, \textcolor{black}{which}
initializes the temperatures of the axial modes of motion to near
the Doppler limit $(\simeq\hbar\gamma_{\mathrm{Be}}/(2k_{{\rm B}})$,
where $\gamma_{\mathrm{Be}}$ is the $^{9}$Be$^{+}$ excited-state
decay rate and $k_{{\rm B}}$ is Boltzmann's constant). We then apply
the EIT cooling beams to $^{24}$Mg$^{+}$ for a cooling dur\textcolor{black}{ation
$t_{c}$. }To determine the final mean motional-state quantum number
$\bar{n}$ of the normal modes, we compare the strength of red and
blue Raman sideband transitions in the $^{9}$Be$^{+}$ ions on the
$|2,1\rangle$ $\rightarrow$ $|1,0\rangle$ \textcolor{black}{transition,
using a pair of 313.220 nm laser beams \cite{Monroe1995,King1998}.}

The $^{9}$Be$^{+}$-$^{24}$Mg$^{+}$ ion pair has two axial motional
modes: a mode where the two ions oscillate in-phase (\textbf{I}) with
frequency $\omega_{\mathrm{I}}/2\pi=2.1$ MHz and an out-of-phase
mode (\textbf{O}) with frequency $\omega_{\mathrm{O}}/2\pi=4.5$ MHz.
The Lamb-Dicke parameters are defined as $\eta=\Delta k_{z}z_{0}$,
where $z_{0}$ is the ground state mode amplitude for the $^{24}$Mg$^{+}$
ion; here, $\eta_{\mathrm{I}}=0.294$ and $\eta_{\mathrm{O}}=0.083$.
The EIT cooling condition $\delta\simeq\omega$ cannot be satisfied
for both modes simultaneously, since the mode frequencies are substantially
different.

We first perform EIT cooling on $^{24}$Mg$^{+}$ for $800$ $\mathrm{\mu}$s,
long enough for the system to reach equilibrium. We set $\Delta/2\pi=96.7$
MHz and $\Omega_{2}/2\pi=12.5$ MHz and scan the value of $\Omega_{1}$
to vary $\delta$ (Eq. (\ref{eq:Stark Shift})).

\begin{figure}[t]
\includegraphics[scale=0.33]{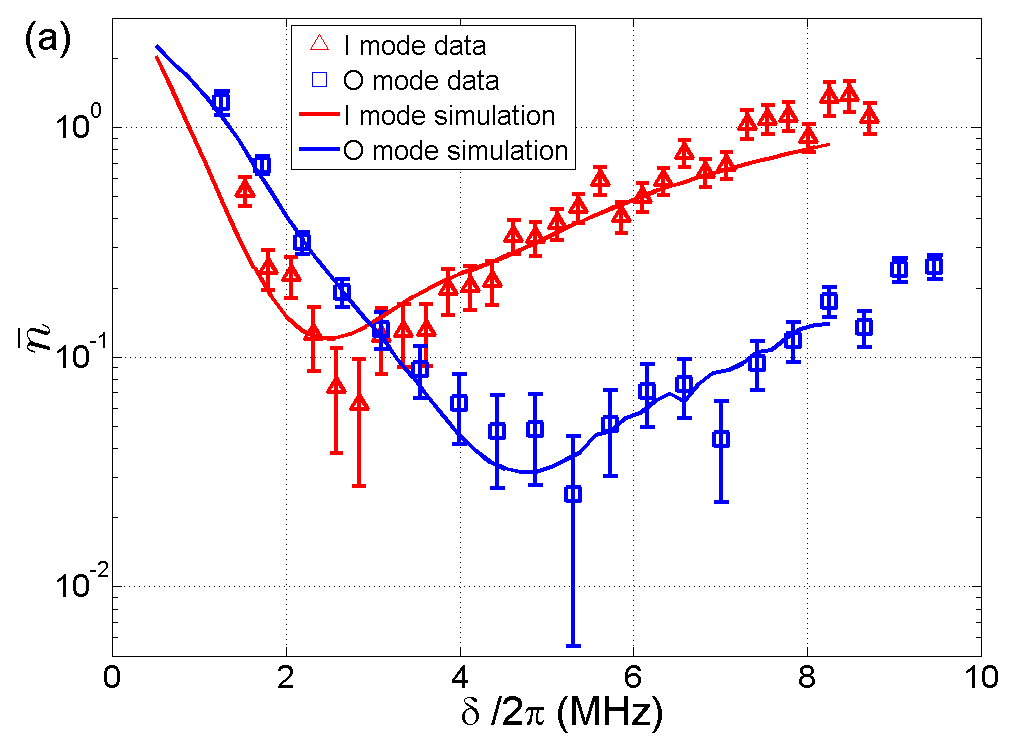}

\includegraphics[scale=0.33]{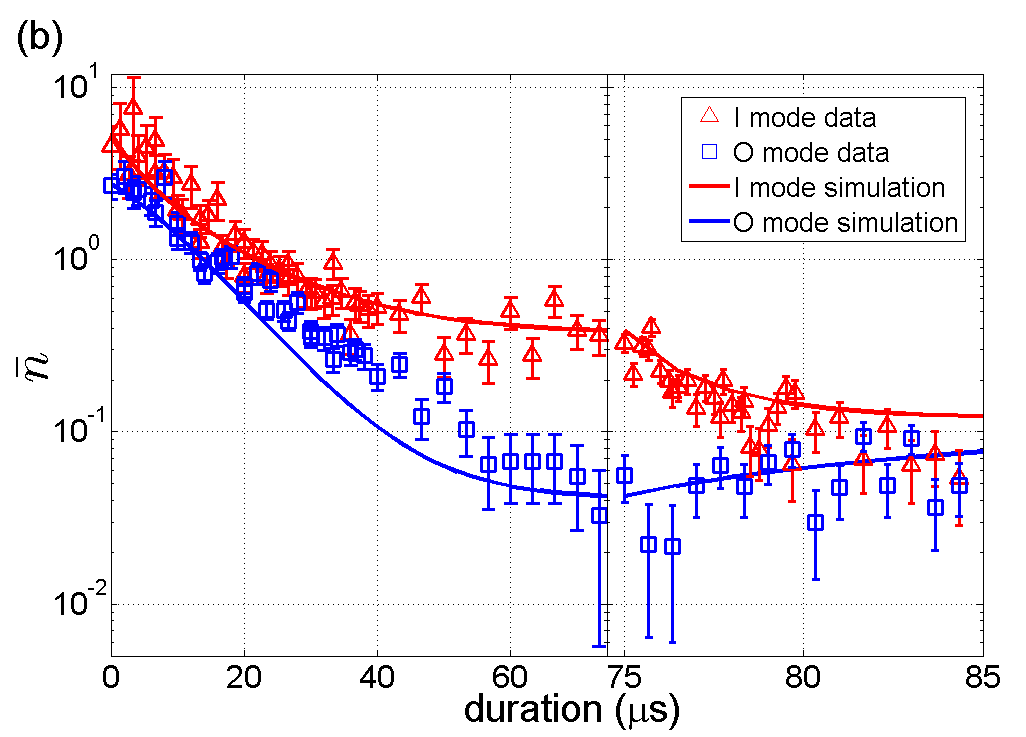}\caption{Mean motional excitation number $\bar{n}$ for the \textbf{I} (red
triangles, $\omega/2\pi$ = 2.1 MHz) and \textbf{O} (blue squares,
$\omega/2\pi$ = 4.5 MHz) axial modes of a $^{9}$Be$^{+}$-$^{24}$Mg$^{+}$
ion pair. (a) $\bar{n}$ after 800 $\mu$s of EIT cooling as a function
of \textcolor{black}{$\delta/2\pi$}. Optimal cooling for each mode
occurs when $\delta$ approximately equals the mode frequency. (b)
$\bar{n}$ plotted as a function of EIT cooling duration $t_{c}$.
From 0 to 74 $\mu$s, \textcolor{black}{$\delta\simeq\omega_{{\rm O}}$}.
From 75 to 85 $\mu$s, $\delta\simeq\omega_{{\rm I}}$ (see text).
In both figures, error bars represent statistical uncertainty of the
sideband amplitude ratios. The solid lines are simulations of the
full dynamics including the $|e'\rangle$ level in $^{24}$Mg$^{+}$\textcolor{black}{,}\textcolor{blue}{{}
}\textcolor{black}{measured ambient heating rates, detuning and beam
intensities. In the simulations we truncated the motion to the first
6 Fock states for both modes for the steady-state simulation in (a)
and to the first 10 (6) Fock states for }\textbf{\textcolor{black}{I}}\textcolor{black}{{}
(}\textbf{\textcolor{black}{O}}\textcolor{black}{) mode for temporal
simulation in (b).} }

\label{fig:TwoionCoolingTimePow}
\end{figure}

The minimum values of $\bar{n}_{\mathrm{I}}$ = 0.08(1) and $\bar{n}_{\mathrm{O}}$
= 0.04(1) are obtained when $\delta$ closely matches the respective
mode frequency, as expected (Fig. \ref{fig:TwoionCoolingTimePow}(a)).
We observe a $\simeq$ 10 \% deviation of the value of $\delta$ needed
for optimum cooling compared to the mode frequency, which can be explained
by additional AC Stark shifts and photon scattering from the $\pi$-polarized
beam that couples $|g_{1}\rangle$ to $|e'\rangle\equiv$ $^{2}$P$_{1/2}\ |m_{J}=-1/2\rangle$
in $^{24}$Mg$^{+}$ (see Fig. \ref{fig:Mgstruc_beam_prof}(a)). We
performed a numerical simulation of the full dynamics including state
$|e'\rangle$ . We also include the effects of heating rates of both
modes, $\dot{\bar{n}}_{\mathrm{I}}$ = 0.38 quanta/ms and $\dot{\bar{n}}_{\mathrm{O}}$
= 0.06 quanta/ms. The average occupation numbers from the simulation
are shown as solid lines in the Fig. \ref{fig:TwoionCoolingTimePow}(a),
and are in good agreement with our experimental results. For the simulations,
we use the treatment of \cite{Morigi2003}, valid in the Lamb-Dicke
regime, adjusted for the relevant modes and mode amplitudes of the
$^{24}$Mg$^{+}$ ions.

To investigate the temporal dynamics of the cooling we set $\delta$
to \textcolor{black}{be near a mode frequency} and measure $\bar{n}$
vs. cooling \textcolor{black}{duration} $t_{\mathrm{c}}$. We first
Doppler-cool both modes with $^{9}$Be$^{+}$ reaching $\bar{n}_{\mathrm{I}}\sim$
5 and $\bar{n}_{\mathrm{O}}\sim$ 2. We find that\textcolor{blue}{{}
}\textcolor{black}{at the experimentally determined optimum values
of $\delta/2\pi$ of 2.55(5) MHz}\textcolor{blue}{{} }\textcolor{black}{and
4.87(5)}\textcolor{blue}{{} }\textcolor{black}{MHz,}\textcolor{blue}{{}
}the $1/e$ cooling time for the \textbf{I} mode is 4(1) $\mu$s and
for the \textbf{O} mode is 15(1) $\mu$s. The faster cooling rate
for the in-phase mode is expected because of its larger $^{24}$Mg$^{+}$
Lamb-Dicke parameter. We can take advantage of the difference in equilibration
times to efficiently cool both modes, as shown in Fig. \ref{fig:TwoionCoolingTimePow}(b).
We first set $\delta\simeq\omega_{\mathrm{O}}$ and apply cooling
for 75 $\mu$s, yielding $\bar{n}_{\mathrm{O}}=0.04(1)$\textcolor{black}{.}
\textcolor{black}{During this stage, the }\textbf{\textcolor{black}{I}}\textcolor{black}{{}
mode is cooled to }$\bar{n}_{\mathrm{I}}=0.36(3)$. We then set $\delta\simeq\omega_{\mathrm{I}}$
and apply the cooling beams for an additional $10$ $\mathrm{\mu}$s,
reaching $\bar{n}_{\mathrm{I}}=0.08(1)$. In this second cooling stage,
the \textbf{\textcolor{black}{O}} mode begins to heat to its equilibrium
value of\textcolor{red}{{} }\textcolor{black}{$\bar{n}=0.19(3)$}, shown
in Fig. \ref{fig:TwoionCoolingTimePow}(a) for this value of $\delta$.
However in 10 $\mu$s, this heating is small, leading to a final value
of $\bar{n}_{\mathrm{O}}=0.07(2)$. Therefore, this two-stage cooling
enables cooling of both modes to near their minimum\textcolor{black}{{}
$\bar{n}$} values in 85 $\mu$s.
\begin{figure}
\includegraphics[scale=0.33]{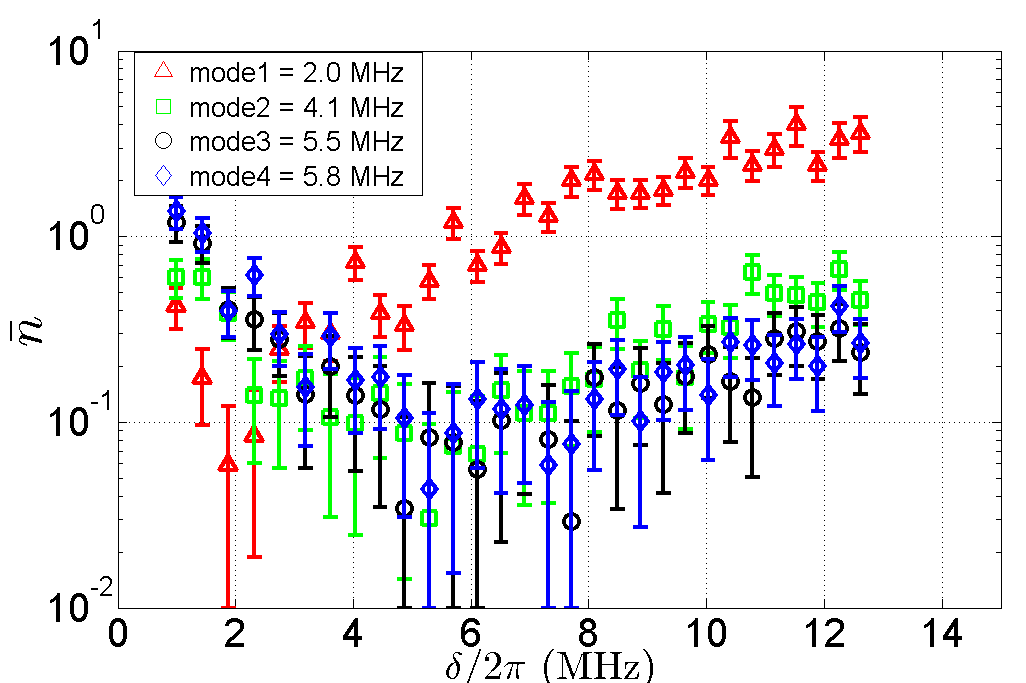}\caption{Minimum \textcolor{black}{$\bar{n}$} values for each of the four
axial modes of a $^{9}$Be$^{+}$-$^{24}$Mg$^{+}$-$^{24}$Mg$^{+}$-$^{9}$Be$^{+}$
ion chain as a function of $\delta/2\pi$ after $800$ $\mu$s of
cooling to ensure steady state. Modes 1 to 4 in the text are labeled as red triangles, green
squares, black circles and blue diamonds, respectively. \label{fig:four ion power scan} }
\end{figure}
\begin{figure}
\includegraphics[scale=0.32]{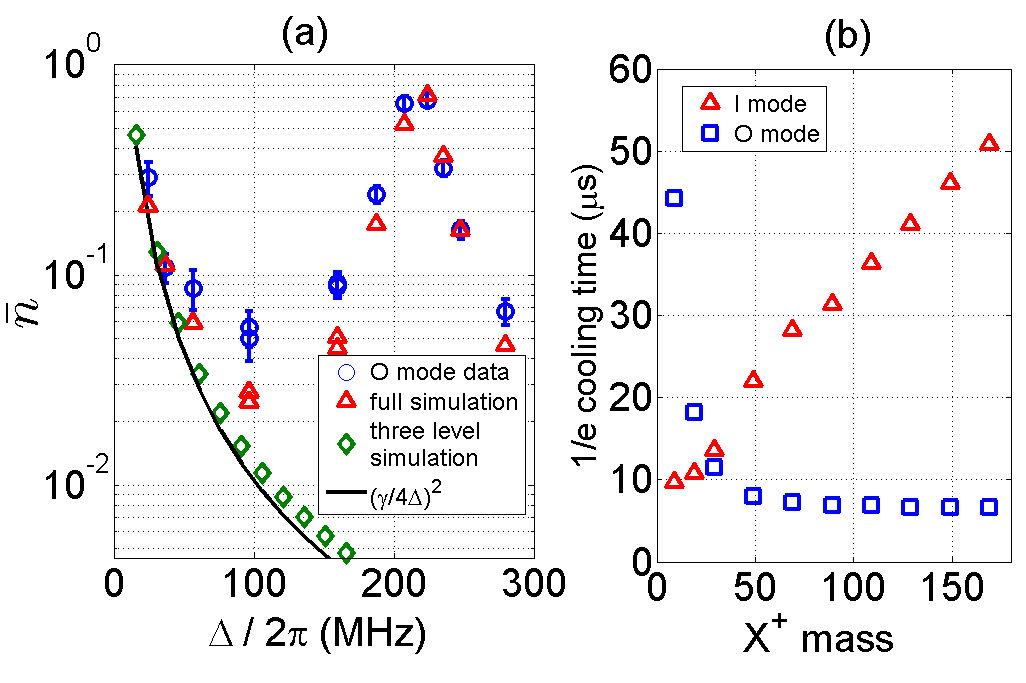}
\caption{(a) Minimum values of $\bar{n}$ for the two-ion \textbf{O} mode vs.
$\Delta/2\pi$. The peak near 223 MHz results from resonant scattering
on the $|g_{1}\rangle\leftrightarrow|e'\rangle$ transition from the
$\pi$ polarized light. Blue circles are the experimental data and
red triangles are simulations based on \cite{Morigi2003}. Green diamonds
are simulations not including\textcolor{blue}{{} }\textcolor{black}{$|e'\rangle$
}and the black solid line shows\textcolor{black}{{} $\bar{n}=(\gamma/4\Delta)^{2}$
\cite{Morigi2000}. (b) Simulation of 1/e cooling time vs. the mass
of ion X$^{+}$ that is sympathetically cooled by }$^{24}$Mg$^{+}$
(axial modes), with fixed trap potential such that $\omega_{{\rm I(O)}}/2\pi=$
2.1 (4.5) MHz when \textcolor{black}{X$^{+}$} is \textcolor{black}{$^{9}$Be$^{+}$}.
Red triangles are for the \textbf{I} mode; blue squares for the \textbf{O}
mode.\textcolor{black}{{}  Optimum values of $\delta$ were chosen for
each mode and ion-mass combination, with $\Omega_{2}/2\pi$ = 5.9
MHz and $\Delta/2\pi$ = 96.7 MHz. (Here $|e'\rangle$ is neglected)}
\label{fig:OOPH-mode-fourth and mass ratio}}
\end{figure}

\textcolor{black}{We also investigate sympathetic EIT cooling for
the four-ion chain $^{9}$Be$^{+}$-$^{24}$Mg$^{+}$-$^{24}$Mg$^{+}$-$^{9}$Be$^{+}$.
}We label the four-ion axial modes \{$1,2,3,4$\}, which have mode
frequencies $\simeq$ $\{2.0,4.1,5.5,5.8\}$ MHz and corresponding
$^{24}$Mg$^{+}$ Lamb-Dicke parameters\textcolor{black}{{} $\{0.21,0.12,0.063,0.089\}$}.
Fig. \ref{fig:four ion power scan} shows the final $\bar{n}$ of
each mode vs. $\delta/2\pi$ after $800$ $\mu$s of cooling to ensure steady state. We set
$\Delta/2\pi$ = 96.7 MHz, $\Omega_{2}/2\pi=$ \textcolor{black}{9.6
}MHz, and scan $\Omega_{1}/2\pi$\textcolor{red}{{} }\textcolor{black}{from
17 to 73 }MHz. The EIT cooling bandwidth is sufficient that modes
2, 3, and 4 can be simultaneously cooled to near their minimum \textcolor{black}{$\bar{n}<0.15$
by setting ${\color{black}\delta/2\pi}$ }= 6.1\textcolor{black}{{}
MHz;} however, at this value, mode 1 is cooled only to\textcolor{black}{{}
$\bar{n}=0.7(1)$. }We therefore again cool in two stages: we apply
40 $\mathrm{\mathrm{\mu}}$s of cooling with \textcolor{black}{${\color{black}\delta/2\pi}$
}=\textcolor{black}{{} }6.1\textcolor{black}{{} MHz} to cool modes 2,
3, and 4 followed by 5 $\mathrm{\mathrm{\mu}}$s of cooling with $\delta/2\pi$
= 2.4 MHz to cool mode 1, reaching $\bar{n}=\{0.11(2),0.20(5),0.14(5),0.18(5)\}$.\textcolor{black}{{} In our experiments, laser beam power of the pi(sigma)-polarized beam ranged between 3 and 10 $\mu$W (3 and 17 $\mu$W). In previous implementations of sequential Raman sideband cooling \cite{Jost09,Home2009,Hanneke2010,Gaebler2012},
cooling of these modes from Doppler temperatures to $\bar{n}\sim0.1$
for each mode required $\sim600\ \mu$s with approximately an order
of magnitude higher laser intensities.}

\textcolor{black}{The $|g_{1}\rangle$ to $|e'\rangle$ transition
frequency is 223.3 MHz higher than that of the $|g_{2}\rangle$ to
$|e\rangle$ transition. Th}us, EIT cooling will be strongly affected
for $\Delta$ near \textcolor{black}{223.3 MHz} due to recoil from
scattering on the \textcolor{black}{$|g_{1}\rangle$ to $|e'\rangle$
transition}. To illustrate the effect, we measure the minimum value
of $\bar{n}$ for cooling the \textbf{O} mode of the $^{9}$Be$^{+}-$$^{24}$Mg$^{+}$
ion pair as a function of the detuning $\Delta$ (Fig. \ref{fig:OOPH-mode-fourth and mass ratio}(a)).
For each value of $\Delta$ we optimize the EIT cooling by varying
$\delta$. The height of this recoil peak depends on the Rabi rate
ratio $\Omega_{2}/\Omega_{1}$ with higher ratios leading to a higher
values of $\bar{n}$. For data of Fig. \ref{fig:OOPH-mode-fourth and mass ratio}(a)
the ratio was held at 0.24. We note\textcolor{black}{{} that for large
detuning $\Delta$, higher laser intensity is needed to maintain values
of $\delta$ near the mode frequencies. }

When $\Delta k_{z}$ is aligned along the trap axis, the motional
modes along the transverse axes are heated by photon recoil. To study
this effect we first \textcolor{black}{cool one of the transverse
modes of a $^{9}$Be$^{+}$-$^{24}$Mg$^{+}$ pair (frequency $\simeq$
4.6 MHz) to near its }ground state with Raman sideband cooling on
$^{9}$Be$^{+}$. We then apply an EIT cooling pulse on th\textcolor{black}{e
}\textbf{O}\textcolor{black}{{} mode, with similar laser beam conditions
as above. After 60 $\mu$s the }\textbf{O}\textcolor{black}{{} mode
is cooled from the Doppler temperature (}$\bar{n}\simeq2$\textcolor{black}{)
to $\bar{n}$ }=\textcolor{black}{{} 0.04(1) while the transverse mode
is heated} from $\bar{n}=0.20(6)$ to $0.9(2)$. Once the \textbf{O}
mode is cooled to near its minimum value, the heating rate of the
transverse mode decreases because the ion becomes approximately trapped
in the dark state for spin and the ground state of axial motion. This
relatively low transverse excitation should cause a negligible error
on a two-qubit gate, which is affected by the transverse modes only
through second-order coupling to the axial mode frequencies\textcolor{black}{{}
\cite{Roos2008}}. Furthermore, Doppler cooling of all modes before
EIT cooling would prohibit any cumulative effect of the heating in
experiments requiring many rounds of sympathetic cooling.

To study the efficiency of EIT sympathetic cooling on other ion species,
such as $^{27}$Al$^{+}$, $^{43}$Ca$^{+}$and $^{171}$Yb$^{+}$
etc., we simulate cooling of an ion pair $^{24}$Mg$^{+}$ - X$^{+}$,
where X$^{+}$ is the sympathetically cooled ion of different mass,
as shown in Fig. \ref{fig:OOPH-mode-fourth and mass ratio}(b). Smaller
differences in ion mass lead to more balanced mode amplitudes and
a reduction in the difference of cooling rates for individual modes.
Large mass imbalances lead to at least one motional mode having a
small $^{24}$Mg$^{+}$ amplitude and thus a long cooling time \cite{Wubbena2012a}.

\label{sec:conclusion}

In summary, we have described sympathetic cooling of $^{9}$Be$^{+}$
ions by EIT cooling of $^{24}$Mg$^{+}$ ions held in the same trap.
We investigate the cooling for both an ion pair and a four-ion chain
crystal that can be used as a configuration for performing entangling
gates between pairs of $^{9}$Be$^{+}$ ions in a scalable architecture
\cite{Jost09,Home2009,Hanneke2010}. \textcolor{black}{By taking advantage
of the different cooling rates for different modes of motion we demonstrated
a two-stage EIT cooling scheme that can bring all modes to near their
minimum excitation level. Compared to previous implementations of
conventional Raman sideband cooling, sympathetic EIT cooling provides
a broad cooling bandwidth, requires less laser power, and is technically
easier to implement. This method may also be useful for sympathetic cooling of molecular ions, for use in quantum logic spectroscopy \cite{schmidt05}, trapped-ion quantum simulation \cite{GDLin2009,Islam2011,Sawyer2012,Islam2012}, strongly-confined neutral atoms \cite{Kampschulte2012}, and nano-mechanical resonators \cite{Xia2009}.}

\begin{acknowledgments}
\textcolor{black}{We thank C. W. Chou and R. Jördens for helpful discussion
on simulation. Also we thank J. J. Bollinger, Y. Colombe, G. Morigi
and T. Rosenband for helpful comments on the manuscript. }This work
was supported by IARPA, ARO contract No. \textcolor{black}{EAO139840},
ONR and the NIST Quantum Information Program. J. P. G. acknowledges
support by NIST through an NRC fellowship. This paper is a contribution
by NIST and is not subject to U.S. copyright.
\end{acknowledgments}


\begin{thebibliography}{10}


\bibitem{CandZ}
J.~I. Cirac and P. Zoller, Phys. Rev. Lett. {\bf 74},  4091  (1995).

\bibitem{Blatt2008}
R. Blatt and D. Wineland, Nature {\bf 453},  1008  (2008).

\bibitem{Blatt2012}
R. Blatt and C.~F. Roos, Nat. Phys. {\bf 8},  277  (2012).

\bibitem{Note1}
Probabilistic schemes for entanglement do not require such strong confinement;
  see C. Monroe, R. Raussendorf, A. Ruthven, K. R. Brown, P. Maunz, L.-M. Duan,
  and J. Kim, arXiv:1208.0391.

\bibitem{Diedrich1989}
F. Diedrich, J.~C. Bergquist, W.~M. Itano, and D.~J. Wineland, Phys. Rev. Lett.  {\bf 62},  403  (1989).

\bibitem{Monroe1995}
C. Monroe, D.~M. Meekhof, B.~E. King, W.~M. Itano, and D.~J. Wineland, Phys.
  Rev. Lett. {\bf 75},  4714  (1995).

\bibitem{Wineland1998}
D.~J. Wineland, C. Monroe, W.~M. Itano, D. Leibfried, B.~E. King, and D.~M.
  Meekhof, J. Res. Natl. Inst. Stand. Technol. {\bf 103},  259  (1998).

\bibitem{kielpinski02}
D. Kielpinski, C. Monroe, and D.~J. Wineland, Nature {\bf 417},  709  (2002).

\bibitem{Gottesman1999}
D. Gottesman and I.~L. Chuang, Nature {\bf 402},  390  (1999).

\bibitem{schmidt05}
P.~O. Schmidt, T. Rosenband, C. Langer, W.~M. Itano, J.~C. Bergquist, and D.~J.  Wineland, Science {\bf 309},  749  (2005).

\bibitem{Jost09}
J.~D. Jost, J.~P. Home, J.~M. Amini, D. Hanneke, R. Ozeri, C. Langer, J.~J.
  Bollinger, D. Leibfried, and D.~J. Wineland, Nature {\bf 459},  683  (2009).

\bibitem{Home2009}
J.~P. Home, D. Hanneke, J.~D. Jost, J.~M. Amini, D. Leibfried, and D.~J.
  Wineland, Science {\bf 325},  1227  (2009).

\bibitem{Hanneke2010}
D. Hanneke, J.~P. Home, J.~D. Jost, J.~M. Amini, D. Leibfried, and D.~J.
  Wineland, Nat. Phys. {\bf 6},  13  (2009).

\bibitem{Jost2010}
J.~D. Jost, Ph.D. thesis, University of Colorado, Boulder, 2010.

\bibitem{Wubbena2012a}
J.~B. W\"{u}bbena, S. Amairi, O. Mandel, and P.~O. Schmidt, Phys. Rev. A {\bf 85},
   043412  (2012).

\bibitem{Gaebler2012}
J.~P. Gaebler, A.~M. Meier, T.~R. Tan, R. Bowler, Y. Lin, D. Hanneke, J.~D.
  Jost, J.~P. Home, E. Knill, D. Leibfried, and D.~J. Wineland, arXiv:1203.3733 and Phys. Rev.
  Lett. {\bf 108},  260503  (2012).

\bibitem{Langer05}
C. Langer, R. Ozeri, J.~D. Jost, J. Chiaverini, B. DeMarco, A. Ben-Kish, R.~B. Blakestad, J. Britton, D.~B. Hume, W.~M. Itano, D. Leibfried, R. Reichle, T.
  Rosenband, T. Schaetz, P.~O. Schmidt, and D.~J. Wineland, Phys. Rev. Lett.
  {\bf 95},  060502  (2005).

\bibitem{Morigi2000}
G. Morigi, J. Eschner, and C.~H. Keitel, Phys. Rev. Lett. {\bf 85},  4458
  (2000).

\bibitem{Morigi2003}
G. Morigi, Phys. Rev. A {\bf 67},  033402  (2003).

\bibitem{Roos2000}
C.~F. Roos, D. Leibfried, A. Mundt, F. Schmidt-Kaler, J. Eschner, and R. Blatt,
  Phys. Rev. Lett. {\bf 85},  5547  (2000).

\bibitem{Schmidt-Kaler2001}
F. Schmidt-Kaler, J. Eschner, G. Morigi, C.~F. Roos, D. Leibfried, A. Mundt,
  and R. Blatt, Appl. Phys. B {\bf 73},  807  (2001).

\bibitem{Webster2005}
S. Webster, Ph.D. thesis, University of Oxford, 2005.

\bibitem{GDLin2009}
G.-D. Lin, S.-L. Zhu, R. Islam, K. Kim, M.-S. Chang, S. Korenblit, C. Monroe,
  and L.-M. Duan, Europhys. Lett. {\bf 86},  60004  (2009).

\bibitem{Islam2011}
R. Islam, E.~E. Edwards, K. Kim, S. Korenblit, C. Noh, H. Carmichael, G.-D.
  Lin, L.-M. Duan, C.-C.~J. Wang, J.~K. Freericks, and C. Monroe, Nat. Commun.
  {\bf 2},  377  (2011).

\bibitem{Sawyer2012}
B.~C. Sawyer, J.~W. Britton, A.~C. Keith, C.-C.~J. Wang, J.~K. Freericks, H. Uys, M.~J. Biercuk, and J.~J. Bollinger, Phys. Rev. Lett. {\bf 108},  213003
  (2012).

\bibitem{Islam2012}
R. Islam, C. Senko, W.~C. Campbell, S. Korenblit, J. Smith, A. Lee, E.~E.
  Edwards, C.-C.~J. Wang, J.~K. Freericks, and C. Monroe, arXiv:1210.0142.



\bibitem{Monroe1995a}
C. Monroe, D.~M. Meekhof, B.~E. King, S.~R. Jefferts, W.~M. Itano, D.~J.
  Wineland, and P. Gould, Phys. Rev. Lett. {\bf 75},  4011  (1995).

\bibitem{King1998}
B.~E. King, C.~S. Wood, C.~J. Myatt, Q.~A. Turchette, D. Leibfried, W.~M.
  Itano, C. Monroe, and D.~J. Wineland, Phys. Rev. Lett. {\bf 81},  1525
  (1998).

\bibitem{Roos2008}
C.~F. Roos, T. Monz, K. Kim, M. Riebe, H. H\"{a}ffner, D.~F.~V. James, and R.
  Blatt, Phys. Rev. A {\bf 77},  040302  (2008).

\bibitem{Kampschulte2012}
T. Kampschulte, W. Alt, S. Manz, M. Martinez~-~Dorantes, R. Reimann, S. Yoon, D. Meschede, M. Bienert, and G. Morigi, arXiv:1212.3814 (2012).

\bibitem{Xia2009}
K. Xia and J. Evers, Phys. Rev. Lett. {\bf 103}, 227203 (2009).



\end{thebibliography}
\end{document}